\documentclass[aps,
twocolumn,
showpacs,pra,amssymb, amsmath]{revtex4}
\usepackage{amsmath,latexsym,amssymb}
\usepackage[dvips]{graphicx}
\usepackage{amsmath}
\usepackage{latexsym}
\usepackage{amssymb}
\usepackage{bm}

\begin{document}

\title{Violation of Rotational Invariance of Local Realistic Models with Two Settings}

\author{Koji Nagata}
\affiliation{ Department of Physics, Korea Advanced Institute of
Science and Technology, Daejeon 305-701, Korea}

\author{Jaewook Ahn}
\affiliation{ Department of Physics, Korea Advanced Institute of
Science and Technology, Daejeon 305-701, Korea}

\date{\today}

\begin{abstract}
We have considered a two-particle Bell experiment to visualize the
conflict between rotational invariance of physical laws and a
specific local realistic theory. The experiment is reproducible by
using a local realistic theory obtained in a two-setting Bell
experiment. The generalized Bell inequality {[J. Phys. A: Math.
Theor. {\bf 40}, 13101 (2007)]}, which is derived under the
assumption that there exists a rotationally invariant local
realistic theory, turns out to disprove such a local realistic model
existing with a two-setting experiment. This implies that such a
model is not rotationally invariant and should, therefore, be ruled
out in some situations.
\end{abstract}

\maketitle

\section{Introduction}
Local and realistic theories assume that physical properties exist
irrespective of whether they are measured and that the result of
measurement pertaining to one system is independent of any other
measurement simultaneously performed on a different system at a
distance. As Bell reported in 1964~\cite{bib:Bell}, certain
inequalities that correlation functions of a local realistic theory
must obey can be violated by quantum mechanics. Bell used the
singlet state to demonstrate this.  Likewise, a certain set of
correlation functions produced by quantum measurements of a single
quantum state can contradict local realistic theories. Since Bell's
work, local realistic theories have been researched
extensively~\cite{bib:Redhead,bib:Peres,bib:jinhyunglee}. Numerous
experiments have shown that Bell inequalities and local realistic
theories are violated~\cite{experiment,experiment2,experiment3}.

In 1982, Fine presented~\cite{bib:fine} the following example: A set
of correlation functions can be described with the property that
they are reproducible by local realistic theories for a system in
two-partite states if and only if the set of correlation functions
satisfies the complete set of (two-setting) Bell inequalities. This
is generalized to a system described by
multipartite~\cite{bib:Zukowski,bib:gene} states in the case where
two dichotomic observables are measured per site. We have,
therefore, obtained the necessary and sufficient condition for a set
of correlation functions to be reproducible by local realistic
theories in the specific case mentioned above.

However, it was shown that such a ``two-setting'' local realistic
model is disqualified if one imposes rotational invariance on local
realistic models with respect to a measurement plane, where one has
more than three spins \cite{Nagata1}. Moreover, in a mixture of
six-qubit Greenberger-Horne-Zeilinger (GHZ)
states~\cite{bib:GHZ,bib:jinhyunglee2}, a generalized Bell
inequality, which is derived under the assumption that there exist
rotationally invariant local realistic models, disproves such a
``two-setting'' local realistic model stronger than a generalized
Bell inequality, which is derived under the assumption that there
exist local realistic models that are rotationally invariant with
respect to a measurement plane~\cite{Nagata2}.

Rotational invariance states that the value of the correlation
function cannot depend on the local coordinate systems used by the
observers. Therefore, we see that such a ``two-setting'' local
realistic model depends on the local coordinate systems used by the
observers in some situation. It was discussed \cite{Nagata3} that
there is a division among the measurement settings, those that admit
local realistic models that are rotationally invariant with respect
to a plane, and those that do not. This is another manifestation of
the underlying contextual nature of local realistic theories of
quantum experiments.

Here, we shall show that such a ``two-setting'' local realistic
model  is disqualified even though one has only two spins if we
impose rotational invariance on local realistic models. This
phenomenon can occur when the system is in a mixed two-qubit state.
We analyze the threshold visibility for two-particle interference to
reveal the disqualification mentioned above. We found that the
threshold visibility is 0.75, which is more stringent than the one
($2(2/\pi)^2\sim 0.81$) reported in Ref.~11. The result implies that
explicit ``two-setting'' local realistic models cannot, in general,
have the property that they are rotationally invariant.

The importance of the result of this paper can be addressed in
conjunction with the convenience to create two-particle
interference. In contrast, it is difficult to create multi-particle
GHZ-type interference. Hence, our result provides a method to
disqualify a rotationally-invariant local realistic theory
experimentally easier than previous discussions in Refs.~11 and 14.

\section{Omnidirectional generalized Bell inequality}

In this section, we shall briefly review the generalized Bell
inequality presented in Ref.~14. Consider two spin-$\frac{1}{2}$
particles, each in a separate laboratory. Let us parameterize the
local settings of the $j$th observer with a unit vector $\vec n_j$
with $j=1,2$. One can introduce the ``Bell'' correlation function,
which is the average of the product of the local results:
\begin{equation}
E(\vec n_1, \vec n_2,) =
\langle r_1(\vec n_1) r_2(\vec n_2) \rangle_{\rm avg},
\end{equation}
where $r_j(\vec n_j)$ is the local result, $\pm 1$, which is
obtained if the measurement direction is set at $\vec n_j$. If the
correlation function admits a rotationally invariant tensor
structure familiar from quantum mechanics, we can introduce the
following form:
\begin{equation}
E(\vec n_1, \vec n_2)
= \hat T \cdot (\vec n_1 \otimes \vec n_2),
\label{et}
\end{equation}
where $\otimes$ denotes the tensor product, $\cdot$ the scalar
product in ${\rm R}^{\rm 3\times 2}$, and $\hat T$ is the
correlation tensor, the elements of which are given by
\begin{equation}
T_{i_1i_2} \equiv
E(\vec x_{1}^{(i_1)},\vec x_{2}^{(i_2)}),
\label{tensor}
\end{equation}
with $\vec x_{j}^{(i_j)}$ being a unit vector of the local
coordinate system of the $j$th observer; $i_j = 1,2,3$ gives the
full set of orthogonal vectors defining the local Cartesian
coordinates. The components of the correlation tensor are
experimentally accessible by measuring the correlation function at
the directions given by the bases vectors in which the tensor is
written. Suppose one knows the values of all $3^2$ components of the
correlation tensor, $T_{i_1i_2}$. Then, with the help of the formula
in Eq.~(\ref{et}), one can compute the value of the correlation
function for all other possible sets of local settings.

We shall derive a necessary condition for the existence of a
rotationally invariant local realistic model of the rotationally
invariant correlation function in Eq.~(\ref{et}). A correlation
function has a rotationally-invariant local realistic model if it
can be written as
\begin{eqnarray}
E_{LR}(\vec{n}_1,\vec{n}_2)=
\int d\lambda \rho(\lambda)
I^{(1)}(\vec{n}_1,\lambda)I^{(2)}(\vec{n}_2,\lambda),
\label{LHVcofun}
\end{eqnarray}
where $\lambda$ denotes some hidden variable, $\rho(\lambda)$ is its
distribution, and $I^{(j)}(\vec{n}_j,\lambda)$ is the predetermined
``hidden'' result of the measurement of all the dichotomic
observables parameterized by any direction of $\vec n_j$. One can
write the observable (unit) vector $\vec n_j$ in a spherical
coordinate system as
\begin{equation}
\vec{n}_j(\theta_j, \phi_j) = \sin \theta_j\cos\phi_j \vec{x}_j^{(1)}
+\sin \theta_j\sin\phi_j \vec{x}_j^{(2)}
+\cos\theta_j \vec{x}_j^{(3)},
\label{vector}
\end{equation}
where $\vec x_j^{(1)}$, $\vec x_j^{(2)}$, and $\vec x_j^{(3)}$
are the Cartesian axes relative to which
spherical angles are measured.

The scalar product of the rotationally invariant local realistic
correlation function, $E_{LR}$ given in Eq.~(\ref{LHVcofun}), with
the rotationally invariant correlation function, $E$ given in
Eq.~(\ref{et}), is bounded by a specific number that depends on
$\hat{T}$. We use the decomposition in Eq.~(\ref{vector}) and
introduce the usual measure $d\Omega_j=\sin\theta_jd\theta_jd\phi_j$
for the system of the $j$th observer. It was proven \cite{Nagata2}
that
\begin{eqnarray}
(E_{LR}, E) & = & \int\!\!d\Omega_1
\int\!\!d\Omega_2
E_{LR}(\theta_1, \phi_1,\theta_2, \phi_2)\nonumber\\
& \times &
E(\theta_1, \phi_1,\theta_2, \phi_2)\leq (2\pi)^2 T_{\max},
\label{Bell-Zineq}
\end{eqnarray}
where $T_{\rm max}$ is the maximal
possible value of the correlation tensor component,
maximized over choices of all possible local settings:
\begin{equation}
T_{\rm max}=\max_{\theta_1, \phi_1,\theta_2, \phi_2}
E(\theta_1, \phi_1,\theta_2, \phi_2).
\label{TE}
\end{equation}
On the other hand, we have
\begin{eqnarray}
(E, E)  =  (4\pi/3)^2 \sum_{i_1,i_2=1}^3T_{i_1i_2}^2.
\label{EEvalue}
\end{eqnarray}
Therefore, the necessary condition for the existence of a
rotationally invariant local realistic model of rotationally
invariant correlations that involve the entire range of settings
reads
\begin{equation}
\max \sum_{i_1,i_2=1,2,3}T_{i_1i_2}^2
\le \left( \frac{3}{2} \right)^2 T_{\max},
\label{3D_NECC}
\end{equation}
where the maximization is taken over
all independent rotations of local coordinate systems
(or equivalently over all possible measurement directions).

\section{Violation of rotational invariance of local realistic models}

Consider two-qubit states:
\begin{eqnarray}
\rho_{a,b}=V|\psi\rangle\langle\psi|+(1-V)\rho_{\rm noise} ~
(0\leq V\leq 1),\label{qubit}
\end{eqnarray}
where $|\psi\rangle$ is the singlet state as
$|\psi\rangle=\frac{1}{\sqrt{2}}(|+^a;-^b\rangle-|-^a;+^b\rangle)$.
$\rho_{\rm noise} = \frac{1}{4} I$ is the random noise admixture.
The value of $V$ can be interpreted as the reduction factor of the
interferometric contrast observed in the two-particle correlation
experiment. The states $| \pm^j \rangle$ are eigenstates of the
$z$-component Pauli observable $\sigma_z^k$ for the $j$th observer.
Here, $a$ and $b$ are the labels of the parties (say Alice and Bob).
One can show that if the observers limit their settings to $\vec
x_j^{(1)} = \hat x_j$, $\vec x_j^{(2)} = \hat y_j$, and $\vec
x_j^{(3)} = \hat z_j$, then one has
\begin{eqnarray}
&&T_{11}=T_{22}=T_{33}=-V,\nonumber\\
&&T_{12}=T_{21}=0,\nonumber\\
&&T_{13}=T_{31}=0,\nonumber\\
&&T_{23}=T_{32}=0.
\end{eqnarray}
Thus, the maximal possible component of the correlation tensor is
equal to $T_{\max} = V$. It is easy to see that
\begin{eqnarray}
\max \sum_{i_1,i_2=1,2,3}T_{i_1i_2}^2=3V^2.
\end{eqnarray}
Hence, the generalized Bell inequality is violated if $V>\frac{3}{4}$.

On the other hand, the set of experimental correlation functions is
described with the property that they are reproducible by
``two-setting'' local realistic theories. See the following
relations along with the arguments in Ref. 8;
\begin{eqnarray}
&&|T_{11}-T_{12}+T_{21}+T_{22}|\leq 2V \leq 2,\nonumber\\
&&|T_{11}+T_{12}-T_{21}+T_{22}|\leq 2V \leq 2,\nonumber\\
&&|T_{11}+T_{12}+T_{21}-T_{22}|=0 \leq 2,\nonumber\\
&&|T_{11}-T_{12}-T_{21}-T_{22}|=0 \leq 2;\\
\nonumber\\
&&|T_{22}-T_{23}+T_{32}+T_{33}|\leq 2V \leq 2,\nonumber\\
&&|T_{22}+T_{23}-T_{32}+T_{33}|\leq 2V \leq 2,\nonumber\\
&&|T_{22}+T_{23}+T_{32}-T_{33}|=0 \leq 2,\nonumber\\
&&|T_{22}-T_{23}-T_{32}-T_{33}|=0 \leq 2;\\
\nonumber\\
&&|T_{11}-T_{13}+T_{31}+T_{33}|\leq 2V \leq 2,\nonumber\\
&&|T_{11}+T_{13}-T_{31}+T_{33}|\leq 2V \leq 2,\nonumber\\
&&|T_{11}+T_{13}+T_{31}-T_{33}|=0 \leq 2,\nonumber\\
&&|T_{11}-T_{13}-T_{31}-T_{33}|=0 \leq 2.
\end{eqnarray}
Therefore, we have
\begin{eqnarray}
\int d\lambda \rho(\lambda) I^{(1)}(\vec
x_1^{(i)},\lambda)I^{(2)}(\vec x_2^{(i)},\lambda)=-V,
\end{eqnarray}
for i = 1, 2, and 3, and
\begin{eqnarray}
\int d\lambda \rho(\lambda) I^{(1)}(\vec
x_1^{(i)},\lambda)I^{(2)}(\vec x_2^{(j)},\lambda)=0,
\end{eqnarray}
for $ i \neq j$.

Please note that the singlet state is $U_1\otimes U_2$  invariant
\cite{WERNER1}. Here, $U_j$ are unitary matrices and $U_1=U_2$.
Hence, for the state $\rho_{a,b}$, we have
\begin{eqnarray}
U_1^{\dagger}\otimes U_2^{\dagger}\rho_{a,b}U_1\otimes U_2=\rho_{a,b}.
\end{eqnarray}
Therefore, one has ``two-setting'' local realistic models  for
values of the correlations for the entire range in space by using
many unitary operations in the form $U_1\otimes U_2$. That is, we
have
\begin{eqnarray}
\int d\lambda \rho(\lambda) I^{(1)}(U_1\vec
x_1^{(i)}U_1^{\dagger},\lambda) I^{(1)}(U_2\vec
x_2^{(i)}U_2^{\dagger},\lambda) \nonumber \\
= -V, \label{Na1}
\end{eqnarray}
for i = 1, 2, and 3, and
\begin{eqnarray}
\int d\lambda \rho(\lambda) I^{(1)}(U_1\vec
x_1^{(i)}U_1^{\dagger},\lambda) I^{(1)}(U_2\vec
x_2^{(j)}U_2^{\dagger},\lambda)\nonumber \\
=0, \label{Na2}
\end{eqnarray}
for $ i \neq j$.

Please note that that these models in Eqs.~(16,17) cannot be ruled
out by any two-setting Bell inequality. The ``two-setting'' local
realistic models in Eqs.~(19,20) are in the following structure
because we used only unitary operations in the form $U_1\otimes
U_2,~(U_1=U_2)$:
\begin{equation}
\int d\lambda \rho(\lambda)
I^{(1)}(\vec{n}_1,\lambda)I^{(2)}(\vec{n}_2,\lambda)
=
\left\{
\begin{array}{ll}
-V, & \vec{n}_1=\vec{n}_2, \\
0, & \vec{n}_1\cdot \vec{n}_2=0.
\end{array}
\right.
\end{equation}
Therefore, no two-setting Bell inequality can rule out the models in
Eqs.~(16,17) because only ``two-setting'' local realistic models
made by only commuting pairs of observables have nonvanishing
values. Nevertheless, despite the fact that there exist
``two-setting'' local realistic models for all directions in
Eqs.~(19,20), and every $U_1\otimes U_2$), these models cannot
construct rotationally invariant local realistic models, and they
are ruled out if $V>\frac{3}{4}$.

Thus, the situation is such that for any value of $V$, one can
construct a ``two-setting'' local realistic model for the values of
the correlation functions for the settings chosen in the experiment
(Eqs.~16,17). One wants to construct an ``omnidirectional'' local
realistic model for the entire range by using ``two-setting'' local
realistic models by using many unitary operations in the form
$U_1\otimes U_2$ and $U_1=U_2$ in (Eqs.~19,20), but these
``two-setting'' models must be consistent with each other, if we
want to construct truly ``omnidirectional'' local realistic models
beyond the $2^2$ settings to which each of them pertains. Our result
clearly indicates that this is impossible for $V > \frac{3}{4}$.
That is, ``two-setting'' models built to reconstruct the $2^2$ data
points, when compared with each other, must be inconsistent;
therefore, they are invalidated. The ``two-setting'' models must
contradict each other. In other words, the explicit models given in
Refs. 8-10 work only for the specific set of settings in the given
experiment, but cannot be rotationally invariant; therefore, they
are ruled out in some situations.

\section{Summary }
In summary, we have shown that  a ``two-setting'' local realistic
model is disqualified, even though one has only two spins, if we
impose rotational invariance on local realistic models. This
phenomenon can occur when the system is in a mixed two-qubit state.
We analyzed the threshold visibility for two-particle interference
to reveal the disqualification mentioned above. We found that the
threshold visibility was 0.75, which is more stringent than the one
($2(2/\pi)^2\sim 0.81$) reported in Ref.~11. The result implies that
explicit ``two-setting'' local realistic models cannot have the
property that they are rotationally invariant.

\section*{Acknowledgments}
This work has been supported by Frontier Basic Research Programs at
Korea Advanced Institute of Science and Technology. K.N. is
supported by a BK21 research professorship.

\end{document}